\documentclass[12pt,preprint]{aastex}

\newcommand\simlt{\lower.5ex\hbox{$\; \buildrel < \over \sim \;$}}
\newcommand\simgt{\lower.5ex\hbox{$\; \buildrel > \over \sim \;$}}

\newcommand{\ms}{\noalign{\vspace{3pt plus2pt minus1pt}}}

\begin{document}

\title{Large-Amplitude, Pair-Creating Oscillations in Pulsar and Black Hole Magnetospheres}
\author{Amir Levinson\altaffilmark{1}, Don Melrose\altaffilmark{2}, Alex Judge\altaffilmark{2}, Qinghuan Luo\altaffilmark{2}}
\altaffiltext{1}{School of Physics \& Astronomy, Tel Aviv University,
Tel Aviv 69978, Israel; Levinson@wise.tau.ac.il}
\altaffiltext{2}{School of Physics, University of Sydney, NSW 2006, Australia}

\begin{abstract}
A time-dependent model for pair creation in a pulsar magnetosphere is developed. It is argued that 
the parallel electric field that develops in a charge-starved region (a gap) of a pulsar 
magnetosphere  oscillates with large amplitude. Electrons and positrons are accelerated 
periodically and the amplitude of the oscillations is assumed large enough to cause creation 
of upgoing and downgoing pairs at different phases of the oscillation. With a charge-starved 
initial condition, we find that the oscillations result in bursts of pair creation in which 
the pair density rises exponentially with time. The pair density saturates at 
$N_\pm\simeq E_{0}^2/( 8\pi m_ec^2\Gamma_{\rm thr})$, where $E_0$ is the parallel electric 
field in the charge-starved initial state, and $\Gamma_{\rm thr}$ is the Lorentz factor for effective pair creation. The frequency  of oscillations following the pair creation burst is given roughly by 
$\omega_{\rm osc}=eE_0/( 8m_ec\Gamma_{\rm thr})$. A positive feedback keeps the system 
stable, such that the average pair creation rate balances the loss rate due to pairs 
escaping the magnetosphere.
\end{abstract}

\section{Introduction}
A key issue in pulsar and black hole electrodynamics is the injection of plasma on open magnetic field lines. In the absence of adequate plasma, referred to as a `gap' where the system is said to be `charge-starved', the rotation of the compact object induces a component of the electric field parallel to magnetic field lines, $E_{||}$, which accelerates any test particle to very high energy. Such `primary' particles  reach energies at which the photons they emit, through curvature radiation (CR) or inverse Compton scattering (ICS), exceed the threshold for effective pair creation, leading to a pair cascade that populates the magnetosphere with `secondary' pairs.  The charge-starved $E_{||}$ is screened once the pair plasma is sufficiently dense to provide the Goldreich-Julian (GJ) charge density.  To maintain the electric charge density close to the GJ value, the plasma must be injected continuously at an  average rate which equals the loss rate due to escape of charged particles along open magnetic lines.     

In a conventional class of pulsar models, (e.g., the reviews by Michel 1991, Beskin, Gurevich and Istomin 1993, Mestel 1998), which we refer to as gap-plus-PFF models, the system is assumed to be in a steady state in the corotating frame. The gap, the region of pair creation and the region where $E_{||}$ is screened are assumed to be spatially separated. Charge starving in an `inner gap' near the stellar surface results in pair creation that is strongly concentrated in a narrow range of heights, referred  to as a pair formation front (PFF) or a pair production front. Reflecting secondaries (positrons if the primaries are electrons) provide a net, quasi-static charge density such that the PFF may be regarded as a thin surfaces with a surface charge density. The resulting additional electrostatic field  screens out the initial $E_\parallel$ above the PFF (e.g., Fawley et al.\ 1977; Arons and Scharlemann 1979; Shibata et al.\ 1998). In some pulsar models pairs are also assumed to be produced in an `outer gap' (Cheng et al.\ 1986), and current conservation can be achieved by the currents through the inner and outer gaps being part of a global circuit (Shibata 1991). More recent models differ quantitatively, rather than qualitatively, from earlier models in that the Lorentz factors of the secondary pairs are smaller (Zhang and Harding 2000, Hibschman and Arons 2000, Arendt and Eilek 2002) and  the pair creation is over a more extensive region, with the heights of the inner and outer gaps perhaps overlapping (Shibata et al.\  2002). Despite extensive studies of such models, it remains unclear how and where the radio emission is generated. One observational estimate of the location of the source of the radio emission (Blaskiewicz,ÊCordes and Wasserman 1991; Gangadhara and Gupta 2001) suggests that it is between the putative inner and outer gaps. Although the possible high-energy emission processes are well understood,  there is no consensus on the location of the high energy emission region, with strong arguments for both the inner gap (Harding and Muslimov 1998) and the outer gap (Romani 1986). The lack of success of pulsar models in explaining the observed emissions (e.g., Beskin 1999) leaves open the possibility that one or more of the basic assumptions in the model may be incorrect. A critical review of the assumptions was given by Michel (2004), who favored a radically different model involving a large Coulomb field associated with charge-separated regions of the magnetosphere (Krause-Polstorff and Michel 1985). 

In this paper we relax the steady-state assumption, arguing that stability of models involving steady-state pair creation at a PFF is questionable. In fact early models  that did not make the steady-state assumption pre-date later models that did. In particular, Sturrock (1971) argued that a steady state is not possible and that the pairs escape in a sequence  of sheets, and Ruderman and Sutherland (1975) invoked pair creation in bursts, called sparking. One specific difficulty with a steady-state model relates to the reflection of secondaries at the PFF. Suppose  $E_\parallel$ accelerates electrons upward, so that it accelerates secondary positrons downward. In a steady-state model most of the secondary positrons are assumed to continue propagating outward, requiring that the screening of $E_\parallel$ above the PFF be sufficient to prevent their reflection. However, when this condition is imposed explicitly (Shibata et al.\ 1988), it requires such implausibly large densities that it seems untenable. If a substantial fraction of the positrons are reflected, such positrons become downward propagating counterparts  of the primary electrons, and by the same sequence of processes can, under some conditions, create a second PFF near the stellar surface (Harding and Muslimov 1998). Reflection of the created electrons at this second PFF would in turn greatly increase the number of primary electrons, leading to a rapidly increasing rate of pair creation. Such a runaway pair creation would not just short out the initial $E_\parallel$, but overshoot it, setting up oscillations with a large amplitude, of order the initial $E_\parallel$. Fine tuning is required in order to have sufficient number of reflected positrons to allow screening, but not too many to cause an overshoot setting up oscillations. The stability of the PFF to oscillations cannot be explored within the framework of conventional gap-plus-PFF models because of the steady-state assumption: the screening is postulated to be independent of time.

Here we explore an alternative model in which $E_\parallel$ is oscillatory, with an amplitude of order the unscreened $E_\parallel$. There is some correspondence between spatially localized phenomena in steady-state gap-plus-PFF models and temporally localized phenomena in an oscillatory model. The inner gap is replaced by an initially charge-starved region that we also refer to as a gap; pair creation associated with the oscillations leads to a burst of pair creation sufficient to screen the initial $E_\parallel$. The spatially localized PFF in a gap-plus-PFF model is replaced by temporally localized pair creation near the phases where $|E_\parallel|$ is maximum. The important new feature in an oscillatory model, is that the electric field is determined primarily by inductive effects, as opposed to electrostatic effects in a gap-plus-PFF model. In fact, we assume that the Goldreich-Julian charge density is small compared with the instantaneous density of electrons or positrons, such that it can be neglected to a first approximation and included through a small asymmetry in the oscillatory model. There is a steady-state current (along a given magnetic flux tube) that is a free parameter in either type of model. However, by hypothesis there is no time-varying current in a steady-state model, whereas the time-varying current is intimately related to the oscillating electric field through the induction equation in the model discussed here. 

The model is introduced in section~2, and numerical results for some illustrative cases are presented in section~3. The model is discussed further in section~4.

\section{Oscillating pair-creating electric field}
In a frame corotating with the star Maxwell's equations can be written in the form,
\begin{eqnarray}
{\bf \nabla}\cdot{\bf E}&=&4\pi(\rho-\rho_{\rm GJ}),
\label{1}
\\
\ms
{\bf \nabla}\cdot{\bf B}&=&0,
\label{2}
\\
\ms
{\bf \nabla}\times{\bf E}&=&-\frac{1}{c}\frac{\partial {\bf B}}{\partial t},
\label{3}
\\
\ms
{\bf \nabla}\times{\bf B}&=&\frac{4\pi}{c}({\bf j}-{\bf j}_R)+\frac{1}{c}
\frac{\partial {\bf E}}{\partial t},
\label{4}
\end{eqnarray}
where $\rho_{\rm GJ}$ is the  GJ charge density, and the term ${\bf j}_R$ is a combination of the fields and their derivatives, and is given explicitly in (Fawley et al.\ 1977).   The electromagnetic fields, current density, and charge density are in the corotating frame.  As shown below, if the difference electric charge density initially satisfies $|\rho-\rho_{\rm GJ}|\sim |\rho_{\rm GJ}|$, then the amplitude of the oscillating current $j$  is of the order of $c\rho_{\rm GJ}$ or larger.  In this case we find $j_R/j\sim \Omega R/c\ll 1$,  
and can therefore neglect the term ${\bf j}_R$ in eq.\ (\ref{4}).

Here we construct a simple 1D, time-dependent model, and consider purely parallel oscillations only.  
We define $s$ to be the distance along  a magnetic field line, and denote by $E_{||}$ and ${\bf E}_{\perp}$ the field-aligned (parallel) and  cross-field (perpendicular) components of the electric field, respectively.  We ignore drift motion (both inertial and ${\bf E}\times{\bf B}$) of charged particles, and assume the motion of pairs to be
restricted to the direction along magnetic field lines.  We further neglect variations of the electric field and charge density across field lines; that is, we assume that  $E_{||}$ and $\rho$ are functions of $s$ and $t$ only.  Equations (\ref{1})--(\ref{4}) then reduce to
\begin{eqnarray}
\frac{\partial E_{||}}{\partial s}&=&4\pi(\rho_e-\rho_{\rm GJ}),
\label{max1}
\\
\ms
\frac{\partial E_{||}}{\partial t}&=&-4\pi(j_{||}-j_{0}),
\label{max2}
\\
\ms
{\bf \nabla}\times{\bf E}&=&0.
\label{max3}
\end{eqnarray}
Here $j_0={\bf b}\cdot({\bf \nabla}\times{\bf B})$ is time independent, where ${\bf b}$ is a unit vector along the magnetic field. Our neglect of variations across magnetic field  lines implies in addition  $\partial j_0/\partial s=0$.  The constant $j_0$ is a free parameter in our model, associated with the global DC current flowing along the corresponding magnetic field line. Equations (\ref{max1}) and (\ref{max2}) are related by virtue of charge conservation. Using ${\bf \nabla}\cdot{\bf j}+\partial\rho_e/\partial t=0$, and the fact that $j_0$ is divergence free, we obtain:
\begin{equation}
{\partial\over\partial t}
\left(\frac{\partial E_{||}}{\partial s}-4\pi\rho_e\right)=0.
\end{equation}
Consequently, Poisson's equation in the form (\ref{max1}) is automatically satisfied, and is needed only to determine the initial electric field $E_{||}(t=0,s)$ for a given choice of initial charge distribution. 

In steady-state models $j_{||}=j_0$ is implicit at all times, so that $\partial E_{||}/\partial t=0$.   Equation (\ref{max1}) is then solved numerically together with the equation of motion for the accelerating pairs, and some prescription  for pair creation is adopted.   In most cases it has been shown that for reasonable pair multiplicities (number of secondary pairs per primary particles) the model predicts some returning positrons, which are usually ignored. However, any change in the number of returning positrons, as necessarily occurs in any perturbation away from the steady state, implies a change in the current density inside the gap from the assumed value $j_0$. According to eq.\ (\ref{max2}), any nonzero difference $j_{||}-j_0$ leads to a temporally varying $E_{||}$.  For a steady-sate model to be stable, the resulting varying $E_{||}$ must eventually be damped so that the steady-state is restored. However, pair creation should lead to an increase in $|j_{||}-j_0|$, until the system overshoots, leading to oscillations in $E_{||}$.  Moreover, it is not at all clear that for arbitrary initial conditions, the pair cascade process can relax to the putative steady-state.   Below we demonstrate that indeed a plausible mode of operation is large amplitude oscillations that control the pair creation rate via a positive feedback, with no indication that the oscillations damp, and hence no indication that the system converges to a steady state.

We adopt a treatment in which the plasma is modeled as a two component fluid, consisting of electrons and positrons.  We find it convenient to introduce the equations in a covariant formalism (Greek indices run over 0,1,2,3, with signature of the metric +2) and then to revert to the one-dimensional (along the parallel axis) form for our specific calculations. The processes incorporated in the model are emission of curvature radiation (CR) photons and one-photon creation of $e^\pm$ pairs on the external magnetic field.   

Let
\begin{equation}
T^{\alpha\beta}_{\pm}=h_{\pm}n_{\pm}U_{\pm}^{\alpha}U_{\pm}^{\beta}
+p_{\pm}g^{\alpha\beta}
\label{TM}
\end{equation}  
denote the  stress energy tensor of electrons ($-$) and positrons ($+$) with  $n_{\pm}$, $U^\alpha_{\pm}$, $p_{\pm}$ and $h_{\pm}$ their proper density, 4-velocity, partial pressure and specific enthalpy, respectively, and let $Q/2$ denote the total creation rate per  unit volume of electrons (or positrons). The continuity equations are
\begin{equation}
\partial_\alpha(n_{\pm}U_{\pm}^{\alpha})=Q/2.
\label{Npm}
\end{equation}
The energy and momentum equations can be expressed as,
\begin{equation}
\partial_\alpha(T^{\alpha\beta}_{\pm})=\pm e n_{\pm}
F^{\beta}{}_{\alpha}U^{\alpha}_{\pm}-S^\beta_{\pm}+Q_{\pm}^\beta,
\label{TMpm}
\end{equation}
where $F^{\alpha\beta}=-F^{\beta\alpha}$ is the Maxwell 4-tensor.  The terms $S^\beta_\pm$ and $Q^\beta$ account, respectively, for radiative losses due to the emission of CR photons, and the change in energy and momentum of the pairs due to pair creation.

The 4-current density is given by 
\begin{equation}
j^{\mu}=e(n_{+}U^{\mu}_{+}-n_{-}U^{\mu}_{-}).
\label{j^u}
\end{equation}
The charge conservation relation $\partial_\mu j^\mu=0$ is implied by eq.\ (\ref{Npm}).

The projection of eq.\ (\ref{TMpm}) on the 4-velocity $U^\beta$ yields  an equation for the change of specific entropy, $\sigma_\pm$, of each fluid:
\begin{equation}
n_\pm T_\pm U_\pm^\alpha\partial_\alpha \sigma_\pm=S^\beta_{\pm}U_{\pm\beta}
-Q_{\pm}^\beta U_{\pm\beta}-h_\pm Q/2,
\label{entropy}
\end{equation}
where $T_\pm$ are the corresponding temperatures.   The first term on the RHS of eq.\ (\ref{entropy}) accounts for the decrease in specific entropy due to radiative losses, and  the last two terms account for the change in the specific entropy resulting from conversion of photons to pairs.

Using eqs (\ref{TM}), (\ref{Npm}), (\ref{TMpm}), and (\ref{entropy}), we obtain the equation of motion of each fluid:
\begin{equation}
n_\pm h_\pm U_\pm^\alpha\partial_\alpha U_\pm^\beta=\pm e n_{\pm}
F^{\beta}{}_{\alpha}U^{\alpha}_{\pm}+(-S^\alpha_{\pm}
+Q_{\pm}^\alpha)\Theta_{\pm\alpha}^\beta-\Theta_\pm^{\alpha\beta}\partial_\alpha p_\pm,
\label{mot}
\end{equation}
where $\Theta_\pm^{\alpha\beta}=g^{\alpha\beta}+U_\pm^\alpha U_\pm^\beta$. 

In order to solve the set of equations we need to make a number of simplifying assumptions, and in doing so we seek the simplest possible model that retains the features that we consider essential. 

We make the following simplifications.  First, under the conditions considered below we anticipate the pressure forces to be much smaller than the electric force.  We therefore neglect the last term on the RHS of eq.\ (\ref{mot}).  Second, we suppose that the specific entropy of each fluid is roughly conserved along streamlines, that is, we take   $U_\pm^\alpha\partial_\alpha \sigma_\pm =0$ in eq.\ (\ref{entropy}).  We emphasize that intense pair creation does not necessarily imply that the entropy per particle must increase. In fact, under the conditions anticipated, the average energy of CR photons is much smaller than the energy of radiating pairs, and so we don't expect the specific entropy to change much during most of the pair creation burst, except for some short initial phase. At any rate, the details should not affect dramatically the general conclusions drawn below.  Third, we ignore any ${\bf E}\times {\bf B}$ drift, and consider only acceleration along magnetic field lines.  Fourth, we adopt simple models for the source terms $S^\beta_{\pm}$ and $Q_{\pm}^\beta$ that are intended to model the effects of CR; we ignore ICS which is mode difficult to model.

We adopt a local coordinate frame in which $F^1{}_0=E_{||}$.  We denote by $U\equiv U^1$ the component of 4-velocity along the magnetic field, by $v$ the corresponding 3-velocity, by $N_\pm=n_\pm\Gamma_\pm$ the pair density as measured in the laboratory frame, and by $s_\pm^1=S_\pm^1/N_\pm$, $q_\pm^1=Q_\pm^1/N_\pm$, and $q_\pm=Q/2N_\pm$, respectively, the radiative drag per particle,  momentum gain per particle due to photon conversion, and pair multiplicity rate. Under the above assumptions equations (\ref{mot}) and (\ref{Npm}) reduce to:
\begin{eqnarray}
\frac{\partial U_\pm}{\partial t}+v_\pm\frac{\partial U_\pm}{\partial s}
&=&\pm\frac{e E_{||} c^2}{h_\pm}-\frac{s^1_\pm}{h_\pm}+\frac{q^1_\pm}{h_\pm}-q_\pm U_\pm,
\label{gap1}
\\
\ms
\frac{\partial N_{\pm}}{\partial t}+\frac{\partial}{\partial s}(N_{\pm}v_\pm)&=&Q/2.
\label{gap2}
\end{eqnarray}
The enthalpies $h_\pm$ may be treated as free parameters of the model (they are assumed to be time independent ).  The maximum Lorentz factor, $\gamma_{\rm max}$, that a test particle can acquire during the initial pair creation burst is limited essentially by radiative drag. To estimate $\gamma_{\rm max}$ during the pair creation burst, we equate the electric force with the CR loss per unit length to obtain: $\gamma_{\rm max}\simeq2\times10^5 E_{||}^{1/4}$.
For a pulsar having $B\le B_c$, where $B_c=m_e^2c^3/e\hbar^2$ is the critical  field strength, the vacuum electric field near the surface is  $E_s\sim \Omega_sB_s R_s\le 5\times 10^8$ V/cm, yielding $\gamma_{\rm max}<10^7$. This implies that the  characteristic energy of CR photons is, $\epsilon_{\gamma\pm}=1.5\gamma_{\rm max}^3\hbar c/\rho_B< 10^5 m_ec^2$, where $\rho_B\simgt R_s$ is the radius of curvature of the magnetic field lines along which the pairs are accelerated. Hence, the energy of newly created pairs is much smaller than the maximum energy $m_ec^2\gamma_{\rm max}$.  Consequently, the fraction of  pairs having maximum Lorentz factor is expected to be rather small during the pair creation burst.  Indeed, as shown below, during the pair creation burst the average Lorentz factor is  limited by mass loading, and is typically well below $\gamma_{\rm max}$.  The amplitude of oscillations of the 4-velocity decreases as the pair density increases, until it reaches a value at which the energy of CR photons is just below the threshold for a single photon annihilation, at which point pair creation ceases.

A detailed derivation of the energy distribution of pairs involves solving the relevant kinetic equations, and a simpler approach is needed for the present purpose.  We assume that the spread in Lorentz factors of particles  in each fluid is small with respect to the bulk Lorentz factor, and take the energy loss  term to be $s^0_\pm=2e^2c\Gamma_\pm^4/3\rho_B^2$ for CR, where $\Gamma_-$ ($\Gamma_+$) is the bulk Lorentz factor of the  electron (positron) fluid.  Under the reasonable assumption that the particles in each fluid are 
distributed isotropically in the fluid rest frame, we obtain 
\begin{equation}
s^1_\pm=s^0_\pm v_\pm=\frac{2e^2c\Gamma_\pm^3U_\pm}{3\rho_B^2}.
\label{s1}
\end{equation}

The characteristic energy of a CR photon is $\epsilon_{\gamma\pm}= 1.5\Gamma_\pm^3\hbar c/\rho_B\simeq 10^{-10.5}\Gamma_\pm^3(\rho_B/10^6 cm)\rm\,eV$.
Provided the energy of the emitted photon satisfies $\epsilon_\gamma\sin\psi>2m_ec^2$, where $\psi$ is the angle between the photon momentum and the direction of the local magnetic field, the photon can decay into an electron-positron pair. The lifetime for this decay is assumed to be short: all photons are assumed to decay instantaneously once the energy of the particle emitting the CR photons exceeds an effective threshold value, $\Gamma_{\rm thr}$.  The number of photons of characteristic energy produced per unit time per  unit volume by the electron fluid is approximately 
$d^2N_\gamma /dtdV\sim N_- s^0_-/\epsilon_{\gamma-}\simeq 10^2 N_-\Gamma_-$, and likewise 
for the positron fluid.  These photons are converted into pairs over a distance 
$\Delta s\ll \rho_B$ if $\Gamma_\pm \gg\Gamma_{\rm thr}$.  Thus, the source terms associated 
with pair creation can be approximated as 
\begin{eqnarray}
Q&=&\alpha_+N_+ + \alpha_-N_-,
\label{Q}
\\
\ms
q_\pm^1&=&(\alpha_+N_+\epsilon_{\gamma+} + \alpha_-N_-\epsilon_{\gamma-})c/2N_\pm,
\label{q1}
\end{eqnarray}
where  $\alpha_\pm=10^2\Gamma_\pm$ for  $\Gamma_\pm>\Gamma_{\rm thr}$, and $\alpha_\pm=0$ otherwise.  The threshold value above which CR photons annihilate inside the gap is a function of position, but for simplicity we take it to be a fixed value (except for one example,
 as explained below), which is a free parameter in our model. The value we adopt in our calculations is $\Gamma_{\rm thr}=10^{6}$.

Once pairs are accelerated to Lorentz factors in excess of $\Gamma_{\rm thr}$,  the pair density rises rapidly.  The bulk Lorentz factor grows until  the right hand side of eq.\ (\ref{gap1}) vanishes, at which point it saturates.  Using  eqs (\ref{s1}) and (\ref{q1}) we find that this happens roughly when $eE_{||}c^2=
h_\pm q_\pm U_\pm$.  When the pair system becomes roughly symmetric, the Lorentz factors of the fluids saturate at a value given by
\begin{equation}
\Gamma_+=\Gamma_-\simeq 10^4\tilde{E}_{||}^{1/2},
\label{G-sat}
\end{equation}
where $\tilde{E}_{||}$ is defined below.   Our interpretation is that the bulk Lorentz factors 
are limited by mass loading.  Specifically, as the bulk Lorentz factors of the accelerating fluids approach this value, the pair creation rate increase to a level at which the electric potential energy goes to create additional pairs that move at the same bulk Lorentz factor rather than accelerating the fluids. 

The above equations can be rendered dimensionless by employing the normalization: $\tilde{E}=eLE/m_ec^2$, $\tilde{N}=N/N_B$, $\tilde{j}_0=j_0/(eN_B c)$, and measuring  velocities in units of $c$, distances in units of $L$ and time in units of $\omega_B^{-1}$. Here $N_B=\Omega_s B_s/2\pi e c$, $\omega_B^2=4\pi e^2 N_B/m_e$, and $L=c/\omega_B$, with  $B_s$ being the magnetic field strength at the stellar surface, and $\Omega_s=2\pi P$ is the angular velocity of the star.  Note that in these units the vacuum electric field near the surface is given as 
\begin{equation}
\tilde{E}_s=eL\Omega_s B_s R_s/m_ec^3=
3\times10^6(B_s/B_c)^{1/2}(P/1\ s^{-1})^{1/2}(R_s/10^6{\rm\,cm}).
\label{E_s}
\end{equation}
In what follows we use this normalization with $B_s=B_c$, $R_s=10^6$ cm, and $P=1$ s$^{-1}$.

\section{Results} 
To obtain our numerical results we integrated eqs (\ref{max2}),  (\ref{j^u}),  (\ref{gap1}), (\ref{gap2}),  (\ref{s1}),  (\ref{Q}), and (\ref{q1}) on a uniform computational grid with $10^4$ cells. In all the examples presented below the inner boundary is at the stellar surface, and the outer boundary above the
pair creation region.  The boundary condition imposed at the outer boundary of the computational domain is the zero-gradient conditions, which minimizes the effect of the boundary condition on the solution. We made some runs with other boundary conditions and found little differences in most cases.

In the first example that we considered we made the simplest assumptions to illustrate the essential features of the model. In this model all quantities are assumed to be initially homogeneous within the gap, and the pair creation rate is also homogeneous with a threshold $\Gamma_{\rm thr}=10^6$.  This is intended to model sparking in a region close to a charge starved stellar surface.  It is readily seen that under these conditions the system remains homogeneous at all times, so that the variations are purely temporal.  An example is shown in fig.~1, where the electric field (normalized to the initial value $E_0=E_{||}(t=0)$), and 4-velocity, normalized to the threshold Lorentz factor ($\Gamma_{\rm thr}=10^6$ in this example) are plotted against dimensionless time, for two different initial conditions.  The upper two panels correspond to initial electric field $\tilde{E}_{0}=-10^4$, and the lower panels to  $\tilde{E}_{0}=-10^6$.  The evolution of the corresponding electron density is exhibited in fig.~2; the positron density equals the electron density. The initial pair densities in both cases were taken to be $\tilde{N}_\pm=10^{-2}$, and the initial velocities $U_{-}=0.95$ and $U_{+}=0$.  The parameter $j_0$ was set equal to the initial current density, viz., $j_0=j_{||}(t=0)=-6.9\times10^{-3}N_Bc$.  We assumed cold fluids: $h_\pm=m_ec^2$.

As clearly seen from fig.~1, the 4-velocity increase rapidly initially until it reaches the value given by eq.\ (\ref{G-sat}), at which point a burst of pair creation commences, as can be seen from fig.~2. This is followed by a phase during which the pair density and, hence, the electric current density, $j_{||}$, rises exponentially while the bulk velocity remains approximately constant.  As $j_{||}$ grows, the rate of change of the electric field, given by eq.\ (\ref{max2}), increases.  The electric field eventually passes through zero and reverses sign leading to a deceleration of the bulk flow.  The system then continues to 
oscillate.  As seen from fig.~1, the amplitude of oscillations of the 4-velocity damps over several periods
until it reaches the pair creation threshold, $\Gamma_{\rm thr}$, at which point the pair cascade process terminates abruptly. 

The sawtooth shape of the electric field oscillations is a consequence of a relativistic effect. The electric current density is proportional to the 3-velocity of the charged  particles, which is approximately constant at $\pm c$ during each cycle, except for a short period during which the particles become non-relativistic briefly, and the current density changes sign abruptly.  To see this more clearly, consider the oscillations
after the pair creation burst ceases and the pair density saturates. Then $N_+\simeq N_-=N_0$ is independent of time, and so $j_{||}=e(N_+v_+-N_-v_-)\simeq 2eN_0v_-$ is also time independent as long as $U_\pm\gg1$.  Eq.\ (\ref{max2}) then implies that $E_{||}$ changes linearly with time during most of the oscillation period.  Specifically,  
\begin{equation}
E_\parallel\approx E_{\rm 0}
\left\{
\begin{array}{ll}
t/T_-
& \hbox{for} -T_-<t<T_-,
\\
(T_-+T_+-t)/T_+
& \hbox{for}\ \  T_-<t<T_-+2T_+,
\end{array}
\right.
\label{sawtooth1}
\end{equation}
where $T_\pm=\omega_B^{-1}\tilde{E}_0/(\tilde{N}\pm\tilde{j}_0)$, with $\tilde{E}_0$ being the amplitude of the electric field oscillations, roughly equals the strength of the initial field, and $\tilde{N}=\tilde{N_+}=\tilde{N_-}$. The sign change of the electric current occurs over a fraction $\Gamma_{\rm thr}^{-1/2}$ of the period.  When $j_0\ne0$ the oscillations are asymmetric in time.  In the
example shown in fig.~1, where $j_0$ is negative, the time it takes the electric field to change from $\tilde{E}_{||}=-\tilde{E}_0$ to $\tilde{E}_{||}=+\tilde{E}_0$ is slightly longer than the time it takes it to change back.  As a consequence the time averaged 4-velocities, $\langle U_-\rangle$ and $\langle U_+\rangle$, do not vanish as in the case of symmetric oscillations.  In fact eqs (\ref{max2}), (\ref{j^u}) and (\ref{gap1}) imply $ec(N_+\langle U_+\rangle-N_-\langle U_-\rangle)=j_0$ as expected. This asymmetry is small in this example, where the DC current $j_0$ is much smaller than the oscillating current $j_{||}$, and is not well resolved in fig.~1. By employing eqs (\ref{max2}) and (\ref{j^u}) and ignoring this small asymmetry we estimate the oscillation period of the saturated system (after pair creation effectively ceases) to be $T=\omega_B^{-1}
(4\tilde{E}_{0}/\tilde{N})$.  Integrating eq.\ (\ref{gap1}) from $U=0$ to $U=U_{\rm max}$, using the above result for the period $T$, and taking the amplitude of the 4-velocity oscillations to be $U_{\rm max}\simeq \Gamma_{\rm thr}$, yields the saturated pair density in terms of the initial electric field strength and the threshold Lorentz factor:
\begin{equation}
N\simeq N_B{\tilde{E_0}^2\over 2\Gamma_{\rm thr}}={E_0^2\over 8\pi m_ec^2\Gamma_{\rm thr}}.
\label{Nsat}
\end{equation}
Substituting the last equation into the above expression for the oscillation period we obtain
\begin{equation}
T=\omega_B^{-1}{8\Gamma_{\rm thr}\over\tilde{E}_0}={8m_ec\over eE_0}\Gamma_{\rm thr}.
\label{period} 
\end{equation}
We find these analytic estimates to be in excellent agreement with the numerical results. Note that there is a positive feedback on the system; any loss of plasma from the gap, as expected in more realistic situations in which the gap is finite, would result in a reduction of the pair density $\tilde{N}$ and a consequent increase of $U_{\rm max}$.  This would lead to regeneration of plasma inside the gap to compensate for the losses.  The average pair creation rate would then be equal to the loss rate.

As a second example we take the initial electric field to be the vacuum field of an aligned rotator near the axis. The vacuum field decreases with height above the stellar surface, and we take this spatial variation into account by adopting a simple model for it. To be more specific, we take $\tilde{E}_{||}(s,t=0)=\tilde{E}_s(1+s/R_s)^{-4}$, with $\tilde{E}_s$ given by eq.\ (\ref{E_s}).  Because the 
magnetic field declines with height (increasing $r$), the absorption length of a single photon at a given energy increases with height, in a manner that depends on the geometry of the magnetic field.  For illustration we take the source terms $Q$ and $q^1$ as in eqs (\ref{Q}) and (\ref{q1}), but with $\alpha_\pm=10^2\Gamma_\pm \exp[-(s/R_s)^2]$ if $\Gamma_\pm>\Gamma_{\rm thr}$ and $\alpha_\pm=0$ otherwise.  In the cases shown in figs~3--5, the initial pair density is 
$\tilde{N}_+(s,t=0)=\tilde{N}_-(s,t=0)=10^{-2}$ for $s\le R_s$ and zero for $s>R_s$, such that initial charge density is $\rho_e(s,t=0)=e(N_+-N_-)=0$.  The initial 4-velocities were chosen such that the initial electric current density is equal to $j_0$. We found that the final results are insensitive to the choice of the initial densities and velocities of electrons and positrons, provided they are much smaller
than the final values. The evolution of the electric field is shown in fig.~3, where the electric field is plotted as a function of distance from the surface (located at $s=0$).  As seen from this time sequence, a wave pattern is formed above the surface, and propagates outwards.  This is a consequence of the
inhomogeneity of the system; since, as discussed above, the frequency of saturated parallel oscillations $T^{-1}$ is roughly proportional to the local electric field strength $E_{||}(s,t=0)$, regions closer to the stellar surface, where the electric field is stronger, oscillate faster, as clearly seen in fig.~3.  The corresponding 4-velocity is displayed in fig.~4: the amplitude approaches the (local) threshold value $\Gamma_{\rm thr}$, as in the first example. The amplitude of the electric field oscillations near the surface become smaller with time.  At $\omega_B t=600$ the electric charge density reduces by a factor of about $10^2$ relative to its value at  $\omega_B t=400$, although the pair density does not change much, cf.\ fig.~5.  As seen from fig.~4, at  $\omega_B t=600$ the oscillations of 4-velocity near the stellar surface become asymmetric, and it appears that both type of charges are convected towards the surface, although we emphasize that 
at this stage a better resolution is needed to resolve the structure near the surface. 

Unfortunately, our numerical resolution is insufficient to follow the system for times longer than shown, and it is not clear what the system evolves towards after very long time. There is an indication in our calculations that after a sufficiently long time the system becomes chaotic.  Physically, we expect that after pair creation effectively ceases, the rate of pair creation falls below the replacement rate needed to overcome escape of pairs, the oscillations break up and a charged-starved configuration is re-established. The evolution described above then repeats. However, we cannot confirm this expectation within the framework of the present model, which does not include the escape of particles explicitly. 

In our final example, we take the initial gap structure to be as in the model considered by Shibata et al.\ (1998).  To be more precise, the initial electric field is given by $\tilde{E}_{||}(s,t=0)=s(1-s/s_0)$ for $s\le s_0$ and $E_{||}=0$ for $s>s_0$, cf.\ fig.~6. In the example shown in fig.~6, $s_0=0.05R_s$. For this choice of parameters the initial electric field is much smaller than the vacuum field given by eq.\ (\ref{E_s}), but still large enough to accelerate pairs well above $\Gamma_{\rm thr}$.  Since $s_0\ll R_s$ the variation of the magnetic field is expected to be rather small and we, therefore, take pair creation to be uniform in the gap.  We find similar results as in the previous examples. Oscillations are set up locally inside the gap, and a wave pattern develops with time by virtue of the non-uniformity of the initial conditions.   Again we are unable to follow the system for times longer than shown.     

\section{Discussion and conclusions}

The preliminary calculations described in the preceding section demonstrate that pair creation in pulsar and black hole magnetospheres is likely to be strongly oscillatory in nature rather than quasi-stationary.  The essential difference between the oscillatory model proposed here and stationary models is in the nature of the response to the $E_\parallel$ that unavoidably develops in a pulsar magnetosphere. In an oscillatory model, this response is essentially inductive, intimately coupled to an oscillating current through the induction equation. In a stationary model $E_\parallel$ is essentially electrostatic and the induction equation is irrelevant by hypothesis. Our analysis shows that a partially unscreened $E_\parallel$ tends to develop large amplitude oscillations whenever the electric current deviates locally from the global DC current flowing in the system.  Such deviations seem unavoidable in realistic situations. This suggests to us that quasi-stationary models are unstable to temporal perturbations, and that the natural state of pair production in a pulsar magnetosphere is oscillatory.
Although in this paper we focus on pulsars, we expect similar behavior in magnetospheres of Kerr black holes.  

Our results show that given an initial charge-starved configuration, the oscillations start with a brief episode during which the pair density rises exponentially, resulting from copious pair creation at phases of the oscillation around the maximum values of $|E_\parallel|$.  The resulting charge (and mass) loading of the oscillating system gives rise to an increase in the frequency of oscillations (at a fixed height), and a decrease in the amplitude of the 4-velocity oscillations. The pair density saturates when the maximum 4-velocity, which is initially well above the threshold required for pair creation, approaches this threshold.  The subsequent oscillations are maintained at a level such that the net creation rate of pairs in the gap balances escape. However, the limitations of our analysis precludes explicit treatment of the escape 
in the simplest form of our model.  
The oscillations in the saturated state are relatively stable, due to a positive feedback whereby any reduction in the pair density leads to an increase in the Lorentz factor of oscillating pairs and, hence, to a net increase in the pair creation rate.  The saturated pair density and corresponding frequency of oscillations depend only on the initial value of the parallel electric field, and the threshold Lorentz factor for pair creation.

In situations where the initial state is not spatially uniform, the frequency of oscillations varies with position in the gap. 
Our calculations show a wave-like pattern developing in the inhomogeneous case, and it appears that the wave-like oscillations 
propagate to regions outside the gap. Our limited resolution does not allow us to follow the evolution of the system for 
long enough time to examine how the oscillations are transmitted to distant regions. 

We regard the calculations presented in this paper as providing a basis for an alternative oscillatory model to the gap-plus-PFF model for the electrodynamics of pulsar magnetospheres. We make a number of major simplifications that need to be relaxed in a more realistic model, but we expect the main features identified here to remain in more realistic models. We comment specifically on three of our simplifying assumptions. %First, we assume that the oscillations are purely temporal and in a more realistic model they should be large-amplitude propagating waves, with both temporal and spatial oscillations. However, provided that phase speed of the waves is greater than $c$, there exists an outward propagating frame in which such superluminous waves are purely temporal oscillations (Rowe 1995), and our calculations should reflect the processes occurring in such frame.
First, the global DC current is a free parameter in our model, and the average charge density is determined from 
initial conditions.  This means that the global structure of the magnetosphere is not addressed within the framework
of our analysis, only the microphysics of charge-starved regions.  
Second, we model the pair creation in a highly idealistic ways: pairs are created only by particles above a specified threshold, $\Gamma_{\rm thr}$, and the pair creation is assumed to be instantaneous. This crude approximation should be adequate to describe the effects of an initial burst of pair creation, starting from a charge-starved configuration, but more accurate approximations are needed to describe the effects of ongoing pair creation once the amplitude of the oscillations approaches a steady state. Also, the simplified model of pair creation is derived assuming that the photons are due to CR, and we do not claim that it is realistic for resonant ICS (Sturner 1995, Luo 1996), which is thought to dominate in some pulsars (Harding and Muslimov 1998).  Third, our assumption that there is no spread in Lorentz factors (the electrons and positrons are cold in their respective instantaneous rest frames) may not be justified under conditions such that the 
spread in Lorentz factors associated with the newly created pairs is comparable to the maximum Lorentz factor due to the oscillating electric field.  This is particularly relevant to in the early phase when a burst of pair creation occurred. Furthermore, our neglect of the initial velocity of newly created pairs during the initial burst is justified only in case of sufficiently strong electric fields, for which the acceleration length is much smaller than the characteristic size of the system.   The inclusion of such effects requires the use of kinetic theory, and this greatly complicates the analysis.

The possible observational implications of the model are not as straightforward as might at first appear. Superficially, one might expect the large-amplitude oscillations to be reflected in temporal structures in the observed emission. For example, one could argue that the period of the oscillations is plausibly of the same order as the time scale of the variations observed in the micro-structure in the radio bursts. However, the emission pattern is strongly dependent on relativistic beaming effects, and the underlying oscillation frequency cannot be readily deconvolved from the Doppler effect and relativistic beaming. A specific prediction of the model arises from the approximate symmetry between upward and downward propagating particles: this suggests that one might observe emission from downward propagating particles above the pole on the opposite side of the compact object to the observer. However, in order to see either high-energy from the opposite hemisphere, its ray path must not intersect the compact object. To see radio emission, not only must it miss the star, but also it must be able to propagate through the region of closed field lines around the star where it can be absorbed or scattered through a variety of processes. The flux of particles impinging on the star must heat the polar cap region, and this can lead to observable thermal X-ray emission. In this context the model is similar to the model of  Harding and Muslimov (1998) with two PFFs.  A careful analysis of the predictions of the model are needed before one can test it with more conventional gap-plus-PFF models, for which predictions are also uncertain. 

The oscillatory model raises some interesting possibilities in connection with the radio emission mechanism. Two mechanisms favored by the model are linear acceleration emission (Melrose 1978; Rowe 1995) associated with the large-amplitude oscillations, and plasma-type emission at the phase where the relative flow of the electrons and positrons is nonrelativistic and Langmuir waves can grow and be converted into escaping radiation (Weatherall 1994). These and other possible radio emission mechanisms require a more detailed discussion than is appropriate here.

AL acknowledges discussions with V. Usov, and 
support by an ISF grant for the Israeli Center for High Energy Astrophysics.

\newpage
\begin{figure}[f1]
\plotone{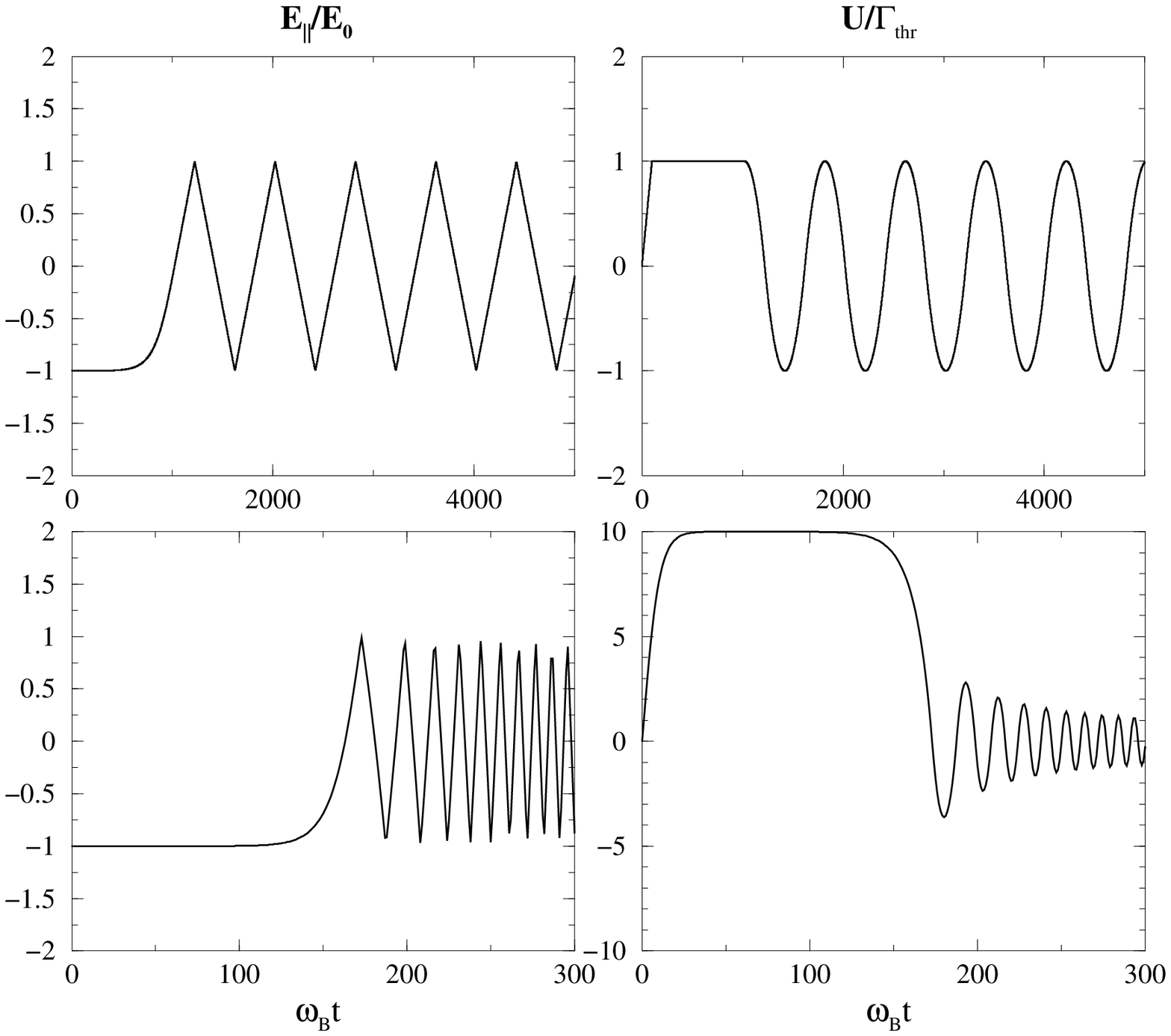}
%\centerline{\epsfxsize=140mm\epsfbox{f1.eps}}
\caption{Evolution of the $E_\parallel$ (left panels) and the 4-velocity (right panels) 
in a large-amplitude oscillation with a uniform initial electric field ${\tilde E}_0=-10^4$ 
(upper panels) and ${\tilde E}_0=-10^6$ (lower panels). The 4-velocity saturates at $U/\Gamma_{\rm th}=1$, 
below which pair creation becomes ineffective, and the system oscillates after the charge screening overshoots.}
\label{f1}
\end{figure}

\newpage
\begin{figure}[f2]
\plotone{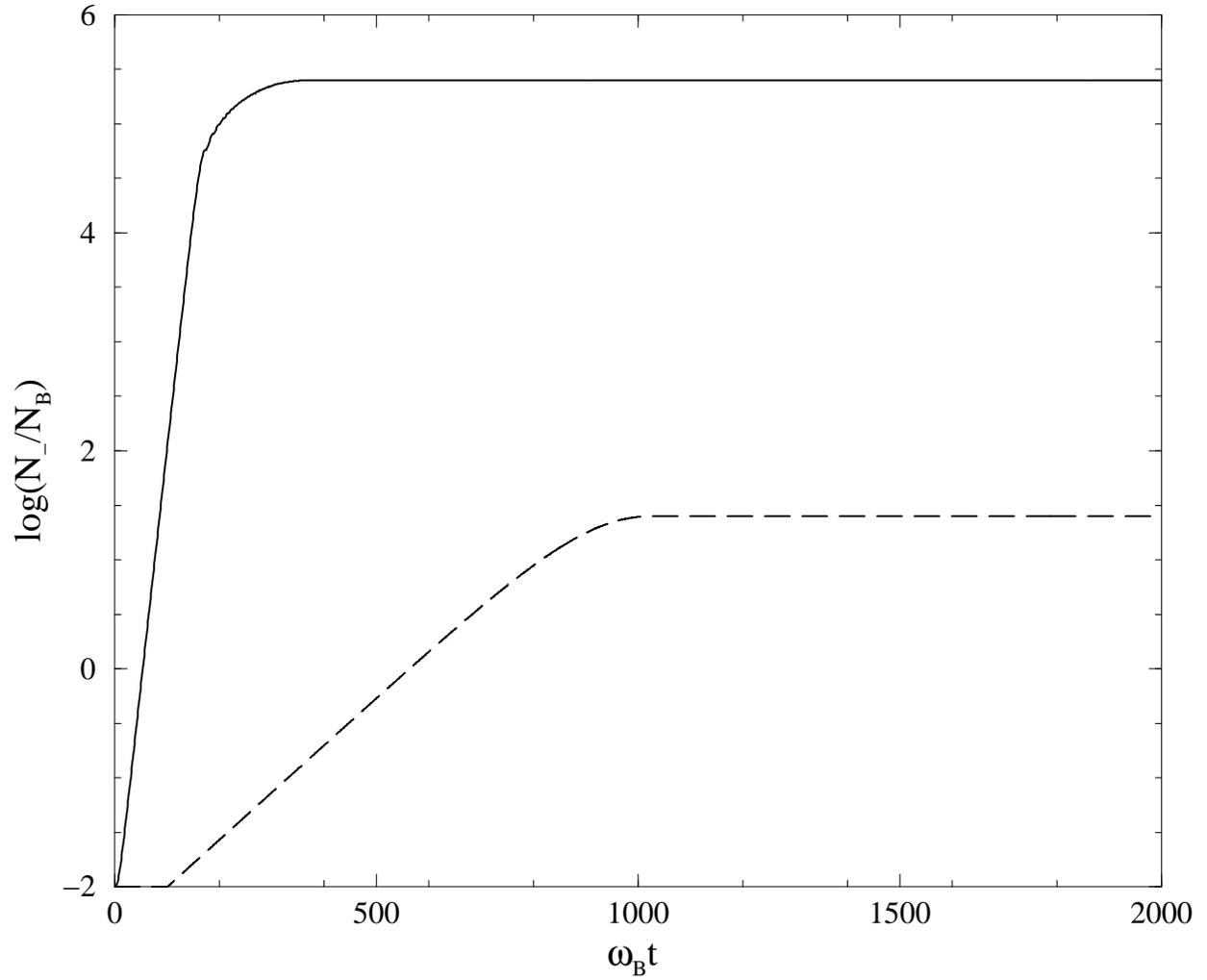}
%\centerline{\epsfxsize=140mm\epsfbox{f2.eps}}
\caption{Evolution of the pair density, showing an exponential increase until the screening overshoots, 
after which it remains roughly constant as the system oscillates. The dashed line corresponds
to the case with ${\tilde E}_0=-10^4$ in fig.~1, and the solid line to  ${\tilde E}_0=-10^6$.}
\label{f2}
\end{figure}

\newpage
\begin{figure}[f3]
\plotone{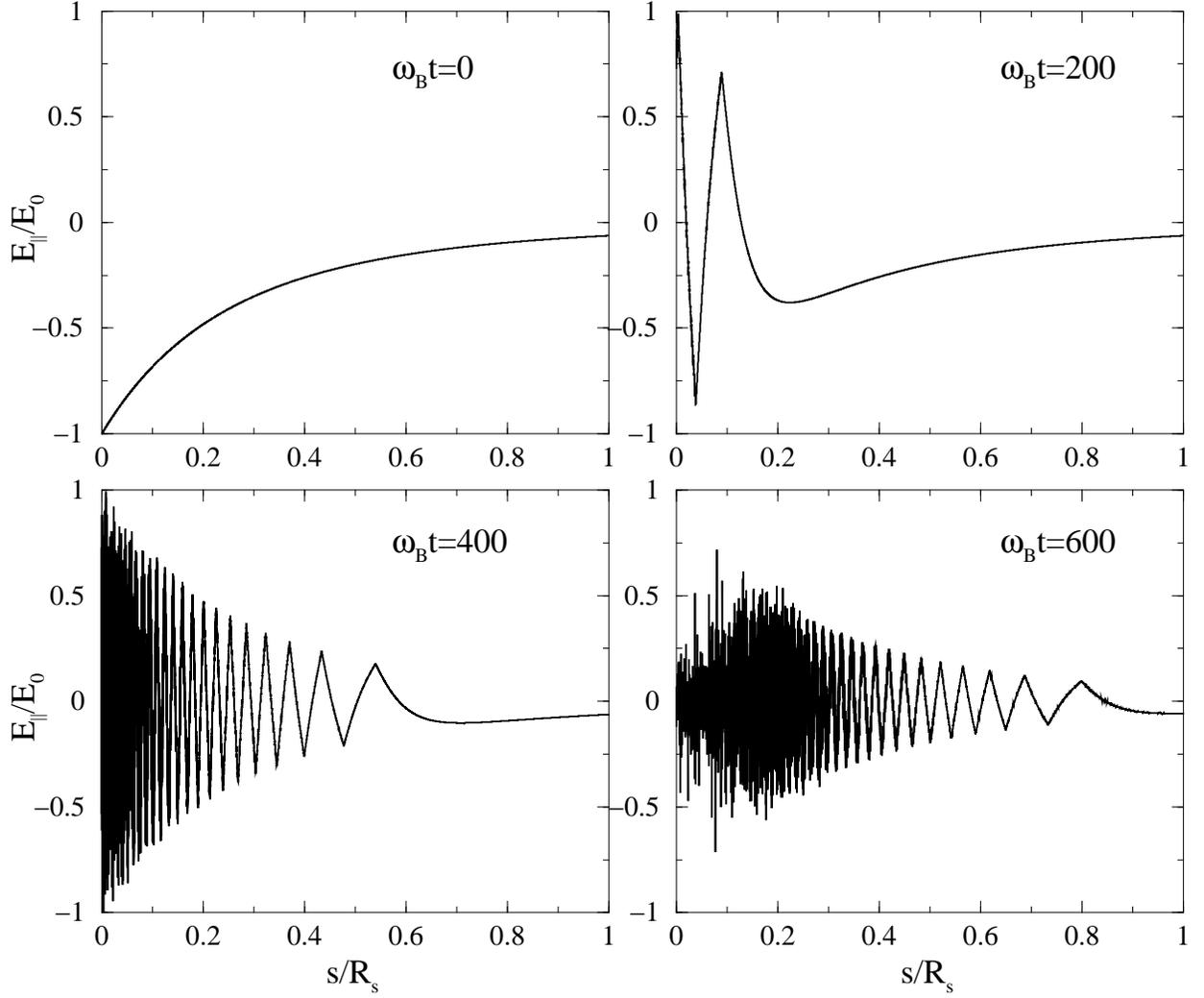}
%\centerline{\epsfxsize=140mm\epsfbox{f3.eps}}
\caption{Evolution of the system starting from a charge-starved condition with 
the vacuum electric field decreasing as illustrated in the top left panel. As time 
increases, oscillations are set up with their frequency and amplitude decreasing with 
height, as shown.}
\label{f3}
\end{figure}

\newpage
\begin{figure}[f4]
%\centerline{\epsfxsize=140mm\epsfbox{f4.eps}
\plotone{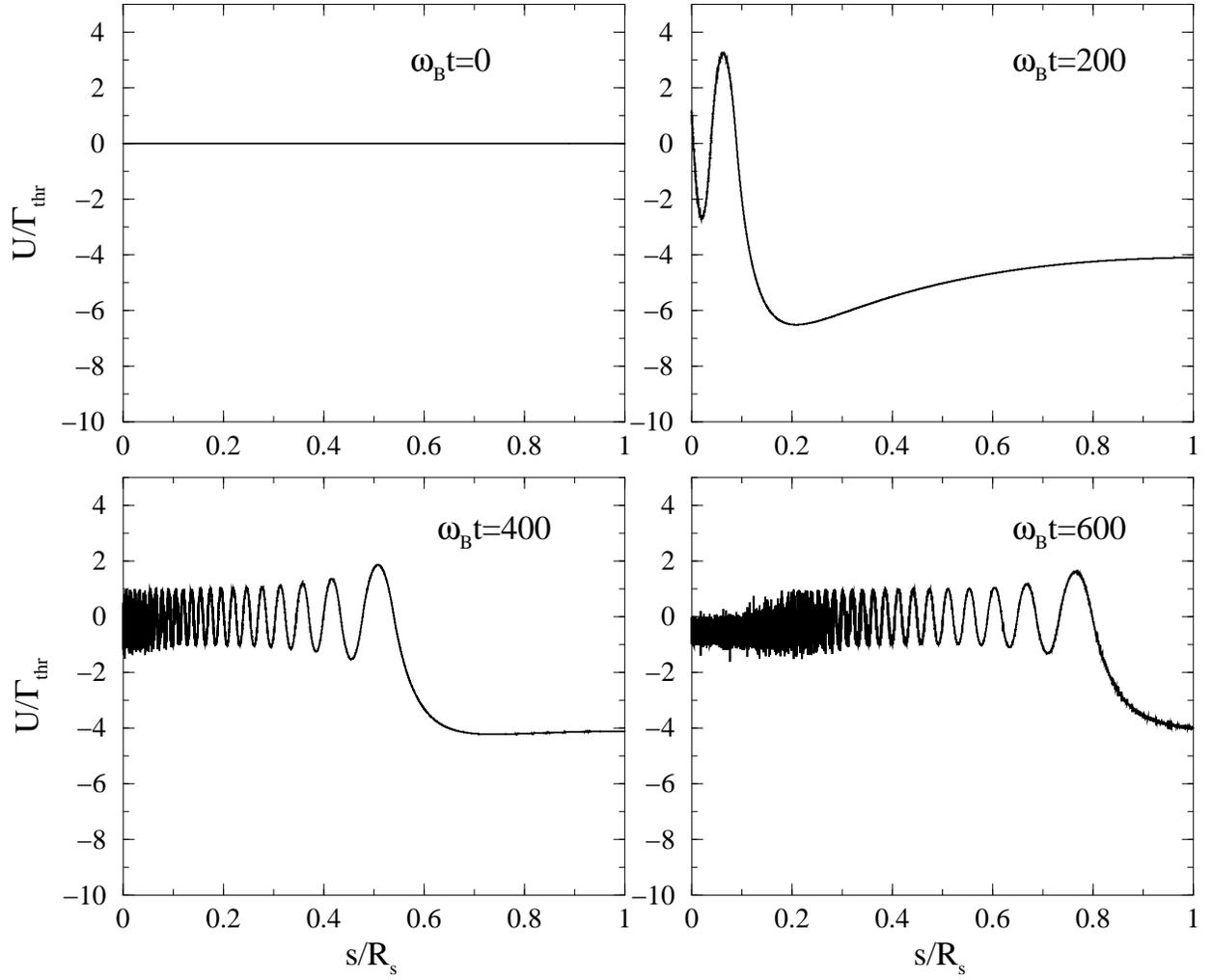}
\caption{ Evolution of the 4-velocity, for the same conditions as in fig 3.}
\label{f4}
\end{figure}

\newpage
\begin{figure}[f5]
%\centerline{\epsfxsize=140mm\epsfbox{f5.eps}
\plotone{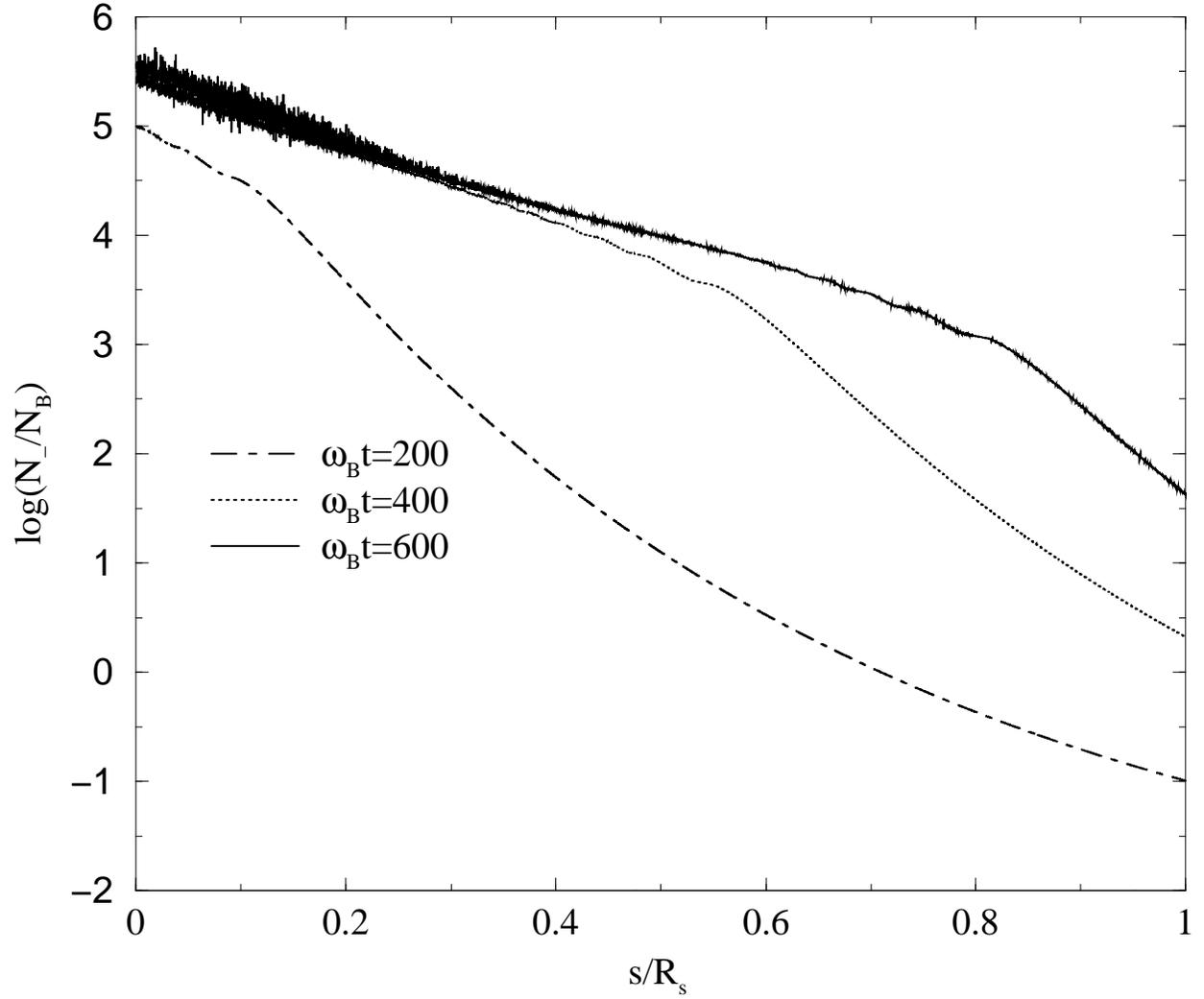}
\caption{The density profile at three different times, computed for the conditions as in fig 3.}
\label{f5}
\end{figure}

\newpage
\begin{figure}[f6]
%\centerline{\epsfxsize=140mm\epsfbox{f5.eps}
\plotone{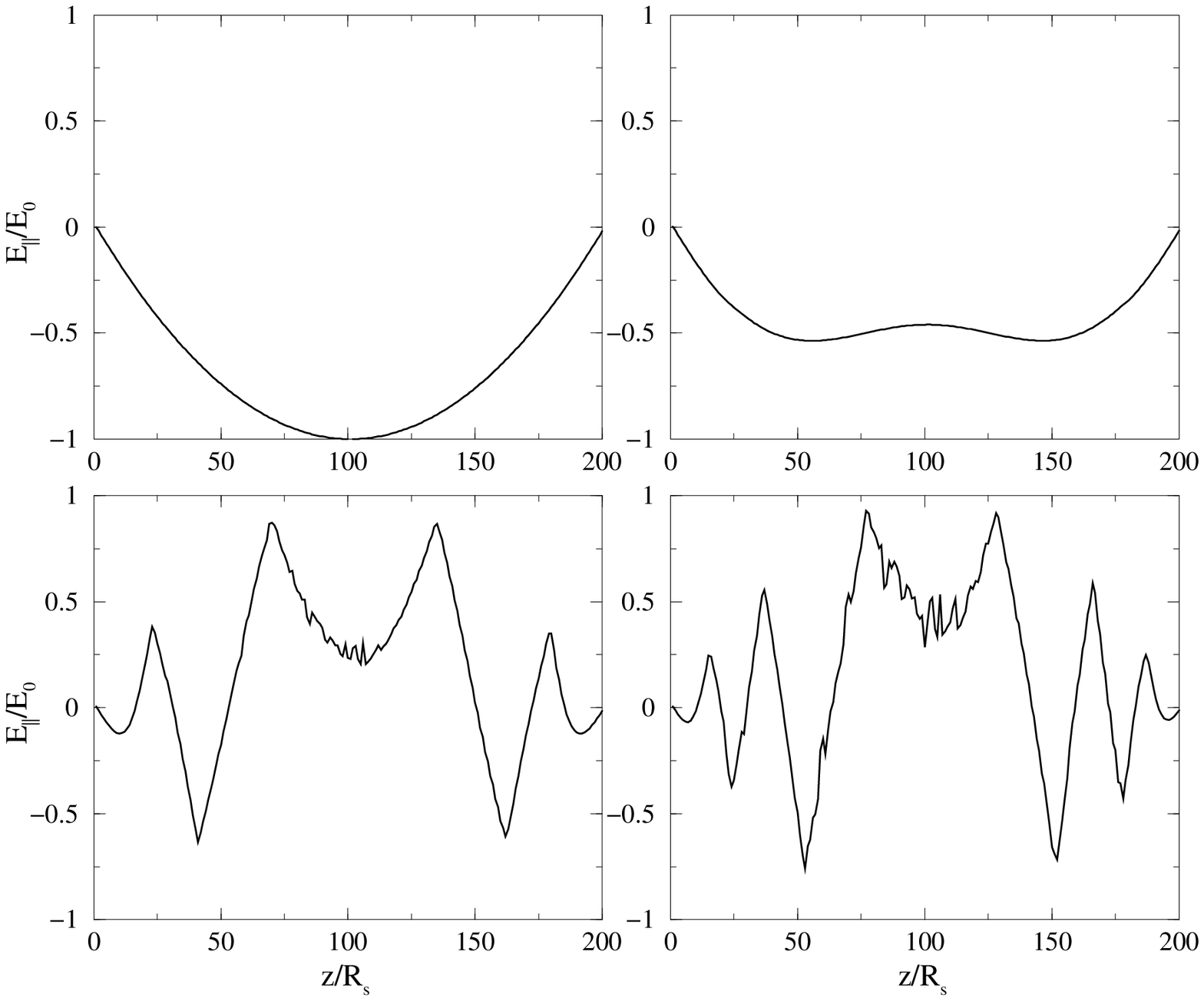}
\caption{Evolution of the system starting from a parabolic parallel electric field,
as shown in the top left panel.}
\label{f6}
\end{figure}


\begin{thebibliography}{9}
\bibitem{} Akhiezer, A. I., and Polovin, R. V. 1956, Sov. Phys. JETP, 3, 696
\bibitem{} Arendt, P. N., Jr and Eilek, J. A. 2002, \apj, 581, 451
\bibitem{} Arons, J. and Scharlemann, E. 1979, \apj, 231, 854
\bibitem{} Beskin, V. S. 1999, Physics-Uspekhi, 42, 1071
\bibitem{} Beskin, V. S., Gurevich, A. V., and Istomin, Ya.N. 1993, Physics of the pulsar magnetosphere, \bibitem{} Blaskiewicz,ÊM., Cordes,ÊJ.ÊM. and Wasserman,ÊI. 1991, \apj, 370, 643
\bibitem{} Cheng, K. S., Ho, C. and Ruderman, M. A. 1986, \apj,  300, 522
\bibitem{} Fawley, W. M., Arons, J. and Scharlemann, E. T. \apj, 217, 227
\bibitem{} Gangadhara, R. T. and Gupta, Y. 2001, \apj,  555, 31
\bibitem{} Harding, A. K., and Muslimov, A. G. 1998, \apj, 508, 328
\bibitem{} Hibschman, J. A., and Arons, J. 2001, \apj,  560, 671
\bibitem{} Krause-Polstorff, l., and Michel, F. C.  1985, \aa, 144, 72
\bibitem{} Luo, Q. 1996, \apj, 468, 338
\bibitem{} Melrose, D. B. 1978, \apj,  225, 557
\bibitem{} Mestel, L. 1998, Stellar magnetism, Oxford University Press
\bibitem{} Michel, F. C.  1991, Theory of neutron star magnetospheres, University of Chicago Press
\bibitem{} Michel, F. C.  2004, Adv.\ Space Res.\ 33, 542
\bibitem{} Romani, R. 1996,  \apj,  470, 469
\bibitem{} Rowe, E. T. 1995, \aap, 296, 275
\bibitem{} Ruderman, M. A., and Sutherland, P. G. 1975, \apj, 196, 51
\bibitem{} Shibata, S. 1991, ApJ, 378, 239
\bibitem{} Shibata, S., Miyazaki, J. and Takahara, F. 1998, MNRAS, 295, L53
\bibitem{} Shibata, S., Miyazaki, J. and Takahara, F. 2002, MNRAS, 336, 233
\bibitem{} Sturner, S. J. 1995, \apj, 446, 292
\bibitem{} Sturrock, P. A. 1971, \apj,  164, 529 
\bibitem{} Usov, V. V. and Melrose, D. B. 1996, \apj,  464, 306
\bibitem{} Weatherall, J. C. 1994, \apj, 428, 261
\bibitem{} Zhang, B. and Harding, A. K. 2000, \apj,  532, 1150
\end{thebibliography}
\end{document}